\documentstyle[epsfig,aps]{revtex}
\input colordvi   
   
\begin{document}   
\thispagestyle{empty}
\title{Probing coherent charmonium photoproduction off light nuclei at 
medium energies}
\author{   
L.~Frankfurt and L.~Gerland\\   
\it School of Physics and Astronomy, Raymond and Beverly Sackler\\   
\it Faculty of Exact Science, Tel Aviv University, Ramat Aviv 69978,\\   
\it Tel Aviv , Israel\\   
M.~Strikman\\   
\it Pennsylvania State University, University Park, Pennsylvania 16802\\   
M.~Zhalov\\   
\it Petersburg Nuclear Physics Institute, Gatchina 188350, Russia}   
\maketitle   
\begin{abstract}
We demonstrate how the elementary amplitudes $\gamma N\to \Psi N$, the
amplitude of the nondiagonal $J/\psi N\Leftrightarrow \psi' N$
transition, and the total $J/\psi N$ and $\psi' N$ cross sections can
be determined from measurements of the coherent $J/\psi$ and $\psi'$
photoproduction off light nuclei at moderate energies. For this purpose we
provide a detailed numerical analysis of the coherent charmonium
photoproduction off silicon within the generalized vector dominance model
(GVDM)  adjusted to account for the physics of charmonium models and color
transparency phenomenon.
\end{abstract}
   
\section{Introduction}   
   
First measurements of quasielastic photoproduction of $J/\psi$ and $\psi'$
mesons off nucleon and nucleus targets at moderate energies were performed
in the seventies at SLAC and Cornell~\cite{{slac2},{slac1},{cornell}}. The
measurement of the A-dependence of $J/\psi $ photoproduction~\cite{slac1}
was used by the authors to extract the $J/\psi$ - nucleon absorption cross
section $\sigma(J/\psi N)=3.5\pm 0.8$~mb, using a semi-classical model for
$J/\psi$ - nucleus interactions. At the same time the estimates of
$\sigma(J/\psi N)$ based on the Vector Dominance Model and the optical
theorem yield $\sigma(J/\psi N)=1.3\pm 0.3$~mb and $\sigma(\psi' N)\approx
0.7 \sigma(J/\psi N)$~\cite{slac2}.  This small value of the $J/\psi N $
cross section comparing to that for $\pi N$ interaction is in variance with
the preQCD ideas on universal hadron-hadron interaction. On the contrary,
these observations can be considered~\cite{FS88} as evidence for the
previously predicted~\cite{{Low:1975sv},{Gunion:iy}} genuine QCD effect of
the dependence of the cross section on the size of the region occupied by
color within a hadron.  Besides, QCD predicts that $\sigma_{\psi' N}$ should
be significantly larger then $\sigma_{J/\psi N}$ because the radius of the
$\psi'$ is two times larger than the radius of the $J/\psi$.  Recently first
data on the hadroproduction of $J/\psi$ and $\psi'$ in pA collisions
confirmed these QCD predictions but large experimental errors and the
complicated theoretical analysis preclude an unambiguous extraction of the
charmonium-nucleon cross sections from the data.  As a result the general
situation with the extraction of these cross sections is still
unsatisfactory - the determined values of $\sigma_{J/\psi N}$ range from 2
mb to 6 mb and, correspondingly, the range of $\sigma_{\psi' N}$ is from 8
mb to almost 20 mb~\cite{vmvogt}. On the other hand the precise values of
these cross sections are urgently needed for the unambiguous interpretation
of the yield of particles with hidden charm produced in the
ultrarelativistic heavy ion collisions (for the review and references see
e.g.~\cite{Kharzeev:1995kz}).
       
Since direct measurements of the charmonium-nucleon cross sections are
impossible, the photoproduction of the $J/\psi$ and the $\psi'$ off nuclei
at coherent moderate energies appears to be one of the most promising tool
to measure these quantities. Really, in the moderate energy domain both the
coherence length and the formation length are smaller or comparable to the
radii of nuclei. So, a photon produces a $c \bar c$ pair which transforms to
a charmonium state inside the nucleus with a noticeable probability and this
meson can interact with the surrounding nucleons. However, it appears that
the standard analysis of the data within the Glauber based Vector Dominance
Model (VDM) which provides a more or less reasonable treatment of the light
vector meson photoproduction at low and moderate photon energies fails in
the case of the charmonium photoproduction.  It was understood that
according to the charmonium models the very small size of the $c\bar c$ pair
which is controlled in the production point by the mass of the $c$-quark
leads to the radical change of the predictions based on the
VDM~\cite{{FS91},{FKS97},{kophuf}}.
    
The aim of this paper is to suggest a new option for the extraction of
charmonium-nucleon amplitudes from measurements of the coherent charmonium
photoproduction off light nuclei at moderate photon energies. In a recent
paper~\cite {fgsz} we demonstrated that large masses of the charmonium
states ($m_V$) and the fluctuations of the strength of the interaction of
different charmonium states with nucleons inside the nuclear medium expected
in QCD lead to significant cross section oscillations in this process.  We
are interested in the photon energy range 15 GeV $\leq E_{\gamma} \leq $ 40
GeV relevant for the planned SLAC experiment~\cite{slac}. In this case the
minimal momenta $q_L\approx {{m_V^2}\over {2\omega}}$ transferred to nucleus
are $0.35 \mbox{ GeV} \geq q_L \geq 0.15\mbox{ GeV}$ and the significant 
oscillations due to the oscillating nuclear form factor are spectacularly 
revealed~\cite{fgsz}
in the energy dependence of the coherent charmonium photoproduction cross
sections.  Another source of oscillations predicted by the QCD color
screening phenomenon is the significant difference of the amplitudes of
$J/\psi N\to J/\psi N$, $\psi' N\to \psi' N$, and $J/\psi N \leftrightarrow
\psi' N$ which leads to oscillations of the rescattering strengths and
noticeably modifies the energy dependence of the cross sections. We suggest
to explore these oscillations as a new method of determining the
elementary photoproduction amplitudes and the amplitudes of $J/\psi$ and
$\psi'$ interactions with nucleons.

Experimentally the suggested approach can be realized at SLAC where    
measuring of the charmonium photoproduction at low energies (15 - 35 GeV)    
is planned~\cite{slac}.
      
\section{Description of the model }    
   
The key feature of the charmonium photoproduction is that the   
$c\bar c$ configuration of the photon wave function is spatially small   
in the production point because its transverse size is controlled by the   
large mass of the $c$-quark.  However, at the photon energies of $15\div 40$ 
GeV
we are interested in, the    charmonium   formation length,
$l_f\approx {{\omega} [{2m_{c}(m_{\psi'}-m_{J/\psi})}]^{-1}}$,     
is still comparable to the internucleon distance and the nucleus radius.    
Hence, the  noticeable probability exists that at 
these moderate   
energies  a
 spatially small configuration would transform  to hadron states before    
a collision with a second nucleon. In this case
using a hadronic basis    
which properly takes into account nonperturbative QCD effects   
seems to be relevant for the calculation of the photoproduction cross   
section   
\begin{eqnarray}   
\sigma_{\gamma A\to \Psi A}(\omega)  =\pi \int \limits_{-\infty}^{t_{min}} 
{\rm d}t   
  {\left |F_{\gamma A\to \Psi A}(t)\right |}^2\quad.   
\label{crosec}   
\end{eqnarray}   
Here $\Psi=J/\psi$ resp. $\psi'$, $-t_{min}=\frac {M_{\Psi }^4} {4\omega^2}$ 
is the
longitudinal momentum transfer in the $\gamma \to \Psi $ transition.  The
photoproduction amplitude $F_{\gamma A\to \Psi A}(t)$ is normalized so that
$\sigma_{\gamma A\to \Psi A}(\omega)$ is in $\mu b$. To calculate this
amplitude the multistep production Glauber approach formulae~\cite{glauber}
were combined with the GVDM~\cite{gvdm} which we adjusted for the charmonium
case in order to for account the color screening effects.  A more detailed
description of this Generalized Glauber Model (GGM) (which we successfully
tested~\cite{fszrho} in the case of the $\rho$-meson coherent
photoproduction) is given in ref.~\cite{fgsz}.  Here we briefly summarize
several essential points.  Since the coherent production of heavy charmonium
states off nuclei at low and moderate energies is suppressed by the target
form factor it is legitimate to build the GVD model restricted to two states
- $J/\psi$ and $\psi'$:
\begin{eqnarray}   
f_{\gamma N \to J/\psi N}=\frac {e} {f_{J/\psi }} f_{J/\psi N\to J/\psi N}+   
\frac {e} {f_{\psi'}}f_{\psi' N\to J/\psi N}\quad,    
\label{gvdm1}   
\\   
f_{\gamma N\to \psi' N}=   
\frac {e} {f_{\psi'}} f_{\psi' N\to \psi' N}+   
\frac {e} {f_{J/\psi }}f_{J/\psi N\to \psi' N}\quad.   
\label{gvdm2}   
\end{eqnarray}   
The charmonium-nucleon coupling constants ${{f_{J/\psi}^2}\over
{4\pi}}=10.5\pm 0.7$, and ${{f_{\psi'}^2}\over {4\pi}}=30.9\pm 2.6$
are determined from the widths of the vector meson decays $\Psi \to e\bar
e$.  Since in the photoproduction processes $c\bar c$ pair is produced
within spatially small configuration one can neglect for a moment the
direct photoproduction amplitude and obtain from
eqs.~(\ref{gvdm1},\ref{gvdm2}) the approximative relations between 
amplitudes
\begin{eqnarray}   
f_{\psi' N\to \psi' N}\approx -{f_{\psi'}\over f_{J/\psi}}   
f_{\psi' N\to J/\psi N}   
\approx {f_{\psi'}^2\over f_{J/\psi}^2}f_{J/\psi N\to J/\psi N}\quad.    
\end{eqnarray}   
Note that a comparatively large absolute value of the nondiagonal amplitude
and the negative value of the ratio of nondiagonal and diagonal amplitudes
follow from the presence of the color screening in the charmonium
photoproduction as dictated by QCD. For a more accurate determination of the
elementary $\psi' N\to \psi' N$ and $J/\psi N\leftrightarrow
\psi' N$ amplitudes within GVDM (for the details see~\cite{fgsz}) we
use as an input in eqs.~(\ref{gvdm1},\ref{gvdm2}) the absolute values of the
forward $\gamma N \to J/\psi N$ cross section, the relation
${{\rm d}\sigma_{\gamma N\to \psi' N}(t_{min})/ {\rm d}t}=   
0.15\cdot {{\rm d}\sigma_{\gamma N\to J/\psi N}(t_{min})/{\rm d}t}$      
and the $J/\psi N$ cross section $\sigma_{J/\psi N}\approx (3.5\pm 0.8)$ mb
found at moderate energies at SLAC~\cite{slac2,slac1}. To show explicitly 
how
uncertainty in the value of $\sigma_{J/\psi N}$ influences on the
photoproduction cross section we used estimates of amplitudes determined
within the corridor of the experimental errors. This is shown by the filled
area in all results of calculations.

\begin{figure}   
\centerline{\hbox{\epsfig{figure=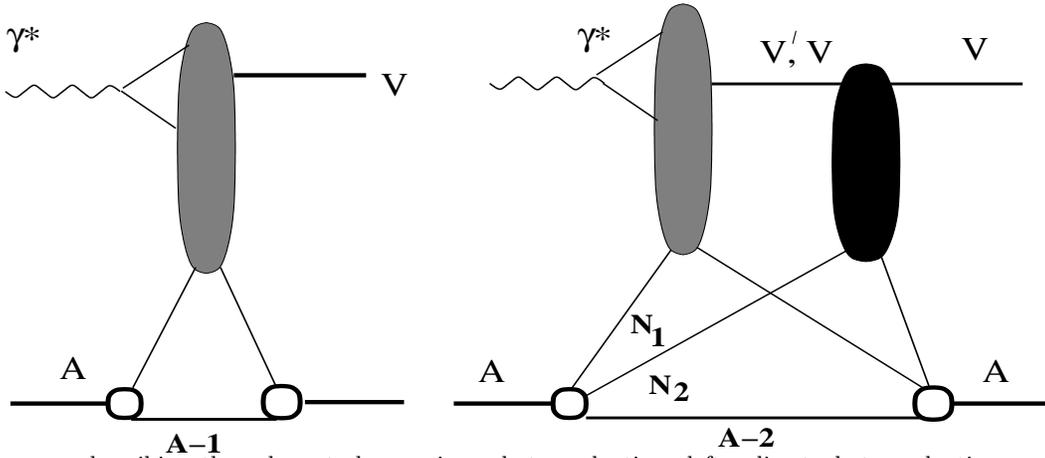,width=14cm,height=6cm}}}
\caption{Diagrams describing the coherent charmonium photoproduction: left -   
direct photoproduction amplitude, right - photoproduction with one 
rescattering    
.}   
\label{diag}   
\end{figure}

We performed calculations within the Generalized Glauber Model (GGM). Note
however, that in the case of coherent photoproduction of charmonium off
light nuclei at moderate energies, the parameter ${\rho \sigma_{\psi N}
R_A}$ is small and one can simplify calculations by taking into account only
one rescattering of the produced heavy meson.  Then only two forward
amplitudes provide the dominating contribution and the coherent charmonium
photoproduction is described by the sum of the diagrams in 
fig.~\ref{diag}. The
first graph is the direct photoproduction amplitude while the second one
accounts for one rescattering of the produced charmonium states within the
nuclear medium.
\nopagebreak   
Since $V,V'=J/\psi,\psi'$ the second diagram includes the
diagonal as well as nondiagonal rescatterings. We want to emphasize that in
the case of nondiagonal transition the effect of charmonium absorption 
enters in the next to leading order.
  
\section{Results and discussion}   
   
Within the Generalized Glauber Model we performed calculations for a wide
range of nuclei trying to choose the optimal nuclei for determining 
the elementary amplitudes from measurements of the coherent photoproduction
of the $J/\psi$ and the $\psi'$. The rates are small for the lightest nuclei
while for sufficiently heavy nuclei multiple rescattering effects become
noticeable.  Hence we found that nuclei with $A=20 \div 40$ are optimal for
our purposes.  Thus in the following we will present the results for 
silicon nuclei. The silicon target is also of interest since one could
envision an active silicon target which would allow to select 
the coherent events in a cleaner way.

\begin{figure}   
\centerline{\hbox{\epsfig{figure=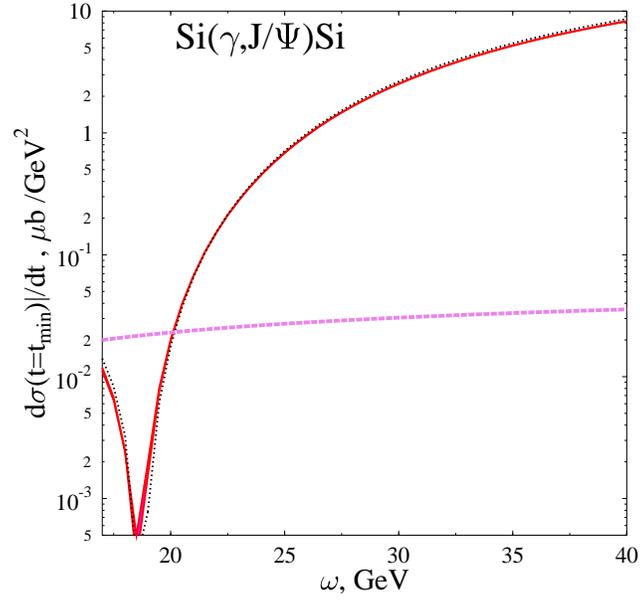,height=10cm}}}
\caption{Energy dependence of the forward coherent $\gamma +Si\to J/\psi
+Si$ photoproduction cross section calculated in the Generalized Glauber
Model compared to the cross sections in the Impulse Approximation (dotted
line).  The forward elementary $\gamma N\to J/\psi N$ cross section is shown
by the dashed line.  
}
\label{sjpsifor}   
\end{figure}

The energy dependence of the  $J/\psi$ cross section is compared 
(fig.~\ref{sjpsifor}) to that obtained within the Impulse Approximation 
where    
\begin{eqnarray}   
{{\rm d}\sigma_{\gamma A\to \Psi A}(t_{min})\over {\rm d}t}=   
{{\rm d}\sigma_{\gamma N\to \Psi N}(t_{min })\over {\rm d}t}\cdot    
{\left |F^{IA}_{\gamma A\to \Psi A}(t_{min}) \right |}^2\quad.   
\label{iacs}   
\end{eqnarray}   
The nuclear density $\rho (\vec b,z) $  is normalized by the    
condition $\int {\rm d}\vec b {\rm d}z\,\rho (\vec b,z)=A$.   
We calculated $\rho (\vec b,z)$ in the Hartree-Fock-Skyrme (HFS)    
model which provided a very good (with an accuracy $\approx 2\%$) 
description    
of the global nuclear properties of spherical nuclei along the periodical    
table from carbon to uranium~\cite{HFS} and the shell momentum distributions    
in the high energy (p,2p)~\cite{p2p} and (e,e'p)~\cite{eep} reactions.   
   
\begin{figure}   
\centerline{\hbox{\epsfig{figure=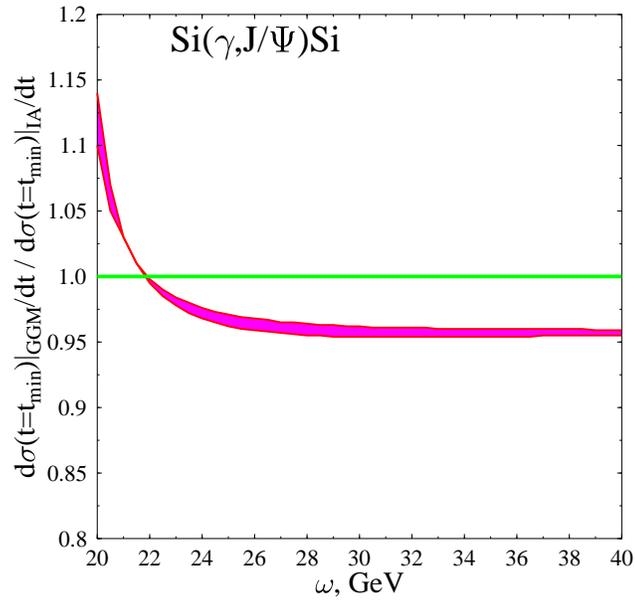,height=10cm}}}
\caption{Energy dependence of the ratio of the forward coherent $\gamma 
+Si\to 
J/\psi +Si$ photoproduction cross section  calculated in the Generalized 
Glauber Model to that in the Impulse Approximation.   
}   
\label{silgia}   
\end{figure}   
    
The distinctive feature of the coherent charmonium photoproduction cross
sections in the kinematics we are interested in is that the dependence of
the cross section on the photon energy is governed by the longitudinal 
nuclear
form factor at a relatively large value of $t_{min}$ which becomes more
important with a decrease of the photon energy. We want to emphasize that 
the
use of a realistic nuclear density in our calculations ensures a
reasonable description of the nuclear form factor in the relevant momentum
transfer range. One can infer from fig.~\ref{silgia} that the cross section
of the $J/\psi$ production off the silicon calculated in GGM practically
coincides with the Impulse Approximation result. The deviation does not
exceed $4\div 5 \%$ and it varies weakly when the rescattering amplitudes 
are
changed within region allowed by the uncertainties of the input values of
the $J/\psi N$ cross section as determined at SLAC~\cite{slac1}.  This 
implies
that one should have an unprecedented accuracy both in the measurements and
in the theoretical analysis in order to extract the $J/\psi N$ elementary
cross section from such measurements. However these measurements can be used
to determine precisely the elementary forward photoproduction cross section.
Really, the contribution of the diagonal $J/\psi N$ rescattering can be
neglected. Then the comparison of the data with the cross section calculated 
in
the Impulse Approximation (eq.~\ref{iacs}) with the nuclear form factor
which is well determined from the high energy electron-nucleus elastic
scattering immediately provides us with the forward elementary $\gamma N\to
J/\psi N$ cross section.  As it is seen from fig.~\ref{sjpsifor}, the low
$t$ cross section of $J/\psi$ production off the silicon in spite of the
strong suppression by the nuclear form factor is considerably larger than
the ${\rm d}\sigma_{\gamma N\to J/\psi N}(t_{min})/{\rm d}t$ (dashed line in
fig.~\ref{sjpsifor}).  Hence, one may significantly improve the accuracy of 
the
determination of ${\rm d}\sigma_{\gamma N\to J/\psi N}(t_{min})/{\rm d}t$.
   
\begin{figure}   
\centerline{\hbox{\epsfig{figure=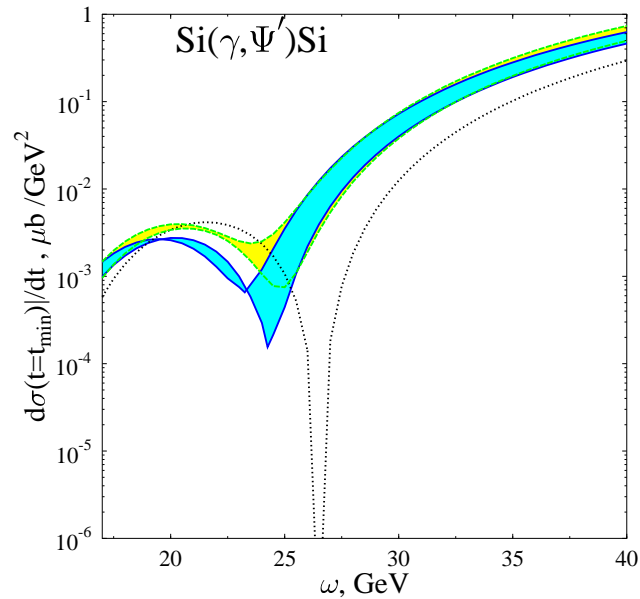,height=10cm}}}
\caption{ The energy dependence of the forward coherent $\gamma +Si\to \psi' 
+Si$ photoproduction cross section calculated in the Generalized Glauber
Model (solid lines, filled blue)  compared to the cross section in the
Impulse Approximation (dotted line) and to the cross section calculated in
GGM without accounting for the diagonal $\psi N\to \psi N$
rescattering (dashed lines, yellow).
}   
\label{spsifor}   
\end{figure}   
   
\begin{figure}   
\centerline{\hbox{\epsfig{figure=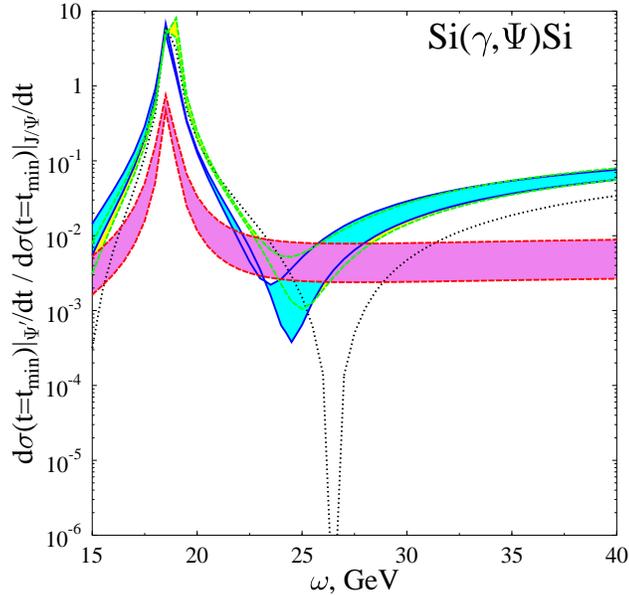,height=10cm}}}
\caption{The energy dependence of the ratio of forward coherent    
$\psi'$-to-$J/\psi$ photoproduction cross sections  calculated   
in the Generalized Glauber Model compared to the ratio in the Impulse    
Approximation (dotted line), to the ratio of cross sections calculated in 
GGM   
without accounting for the diagonal rescatterings    
 (dashed line, yellow filled) and to the ratio without accounting for the 
direct $\psi'$ photoproduction (dashed lines, violet filled) .    
}   
\label{silrat}   
\end{figure}   
   
The situation is qualitatively different for the coherent $\psi'$
photoproduction off silicon (fig.~\ref{spsifor}). The cross section
calculated within GGM significantly exceeds the Impulse Approximation cross
section. This is because the suppression due to the diagonal rescatterings 
is
negligible as compared to the enhancement due to the contribution of the
two-step nondiagonal $\gamma A\to J/\psi A\to \psi' A$ process.  The range
of energy in the vicinity of the form factor minimum for the forward
$\psi'$ production is especially interesting. In this case the main
contribution to the cross section (fig.~\ref{silrat}) originates from the
nondiagonal rescattering.  In particular, at $\omega\approx 0.13R_A
m_{\psi'}^2$ one can extract from the data the nondiagonal elementary
$J/\psi N\to \psi' N$ amplitude by measuring the forward relative $\psi'$
-to-$J/\psi$ yield. In such a ratio all other inputs, namely, the nuclear
density and the elementary $\gamma N\to J/\psi N$ amplitude are already
fixed. Moreover, since they enter both in the numerator and denominator, the
major uncertainties are canceled out.

In the discussed energy range the soft $\Psi N$ rescattering processes
within the nuclear medium are characterized by a rather weak energy
dependence, like $s^{0.08}$, which can be neglected to a very good accuracy
in the ratio of the forward $\psi' $ and $J/\psi$ photoproduction cross
sections. Hence the energy dependence of this ratio (fig.~\ref{silrat})
should originate primarily due to the contribution of the direct $\psi'$
production. Thus one would be able to determine the $\gamma N\to \psi' N$
amplitude by measuring the relative $\psi'$-to-$J/\psi$ yield.

We want to emphasize that the suggested procedure for extracting the
nondiagonal amplitude and amplitudes of direct $J/\psi$ and $\psi'$
photoproduction from the nuclear measurements is practically model
independent.

The task of determining the diagonal $J/\psi N\to J/\psi N$ and $\psi' N\to 
\psi' N$ amplitudes which are relevant for the suppression of the charmonium 
yield in heavy ion collisions is much more complicated.

If one would naively use the VDM, neglecting nondiagonal 
transitions and using values of $\sigma_{tot}(J/\psi N)$ based on the 
SLAC data~\cite{slac1} one would predict a rather significant
suppression of the  $J/\psi$  yield: 
 $\approx 10\%\div 15\%$ for light nuclei, and 
 $\approx 30\%\div 40\%$ for heavy nuclei. However,    
we find~\cite{fgsz} a strong compensation of the suppression due to the
contribution of the nondiagonal transitions. As a result we find that
overall the suppression does not exceed $5\%\div 10\%$ for all nuclei along
the periodical table.  Hence, an extraction of the diagonal amplitudes from
the measured cross sections would require a comparison of high precision
data with very accurate theoretical calculations including the nondiagonal
transitions. Thus one may try another strategy. If the elementary $\gamma
N\to J/\psi N$, and $\gamma N\to \psi' N$ photoproduction amplitudes,
as well as the nondiagonal amplitude $J/\psi N\leftrightarrow \psi' N$
would be reliably determined from moderate energy data it would be
possible to determine the imaginary parts of the forward diagonal amplitudes
from the GVDM equations~(\ref{gvdm1},\ref{gvdm2}) reversing the procedure
which we used in section 2, namely
\begin{eqnarray}   
\Im f_{J/\psi N\to J/\psi N}= \frac {f_{J/\psi }} {e} \Im  
f_{\gamma N \to J/\psi N}-   
\frac {f_{J/\psi }} {f_{\psi'}} \Im f_{\psi' N\to J/\psi N}\quad,    
\label{gvdm3}   
\\   
\Im f_{\psi' N\to \psi' N}=   
\frac {f_{\psi'}} {e} \Im f_{\gamma N\to \psi' N} -   
\frac {f_{\psi'}} {f_{J/\psi }} \Im f_{J/\psi N\to \psi' N}\quad.   
\label{gvdm4}   
\end{eqnarray}   
The main limitation of the suggested procedure is the restriction of the
hadronic basis to the two lowest 1S, 2S charmonium states with the photon
quantum numbers: $J/\psi$ , $\psi'$. This question was discussed 
in~\cite{fgsz}. The
disregard by the closest higher charmonium state $\psi(3770)$ can change
the estimate of the $J/\psi$  N amplitude by $\approx 10\%$.  Influence of 
the
higher mass resonances is expected to be even weaker - the constants
$1/f_{V}$ relevant for the transition of a photon to a charmonium state 
$V$ rapidly decrease with the resonance mass. This is because the radius of 
a bound state, $r_V$, is increasing with the mass of the resonance and
therefore the probability of the small size configuration being $\propto
1/r_{V}^3$ is decreasing with an increase of mass (for fixed S,L).  
Besides, the asymptotical freedom in QCD dictates decreasing of the coupling
constant relevant for the behavior of the charmonium wave function at small
relative distances.  Experimentally one finds from the data on the leptonic
decay widths that $1/f_{V}$ drops by a large factor with increasing mass. 
An additional suppression arises due to the weakening of the soft
exclusive nondiagonal $V N\leftrightarrow V' N$ amplitudes between
states with the different number of nodes. Hence, determining the
imaginary parts of diagonal rescattering amplitudes from the GVDM equations
seems to be possible.  Since in the medium energy domain the energy
dependence of soft rescattering amplitudes is well reproduced by a
factor $s^{0.08}$ the real parts can be found using the well known
Gribov-Migdal relation
\begin{equation}   
{\Re f_{\Psi N\to  \Psi N}}   
={s\pi\over 2}{{\partial \over \partial    
\ln{s}}{\Im f_{\Psi N\to \Psi N}\over s}}\quad.   
\end{equation}   
   
Using the suggested procedure one will be able to determine all elementary
amplitudes with a reasonable precision by measuring the photoproduction of
$J/\psi$  and $\psi'$ off the light nucleus at the medium photon energies.  
The
cross check of this approach would be a comparison of the cross sections
calculated within GGM with parameters fixed in the analysis of the light
nuclei with the experimental cross sections measured in photoproduction of
charmonium off heavy nuclei as well as in the quasielastic processes.
   
\section{Conclusions}   

We used the Generalized Glauber Model which combines the multistep Glauber
approach and the Generalized Vector Dominance Model to calculate the
coherent charmonia photoproduction off light nuclei at medium energies
corresponding to the kinematics of the planned E160 SLAC experiment. We
demonstrate that the nondiagonal amplitudes which follow from the
significant QCD color fluctuations within hadrons give a significant
contribution to the coherent cross section of the $\psi'$ photoproduction
off light nuclei in this energy domain via the two step production
mechanism, while the production of the $J/\psi$ is well described by the
Impulse Approximation.  We show how one can determine the elementary
amplitudes of the charmonia photoproduction as well as the genuine $J/\psi
N$, $\psi' N$ cross sections and the nondiagonal amplitude from measurements
in this domain.

\vspace*{1cm}
{\bf Acknowledgement:}\\   
This work was supported in part by GIF, DOE. L.G. thanks the Minerva   
Foundation for support.

\end{document}